\documentclass[twocolumn]{aastex7}
\usepackage{amsmath}
\usepackage{booktabs}   
\usepackage{graphicx}
\usepackage{subcaption}  
\usepackage{tabularx}   
\usepackage{array}      
\usepackage{threeparttable}
\usepackage{multirow}
\usepackage{bibentry}
\usepackage{subcaption}
\usepackage{threeparttable}

\begin{document}
\title{Measurement of the Orbital Parameters, Spin and Spectral Evolution During the Main High State of Her X-1 with Insight-HXMT}
\author[orcid=0009-0009-1454-9073]{Wen Yang}
\affiliation{Department of Astronomy, School of Physics and Technology, Wuhan University, Wuhan 430072, China}
\email{yangwen63@whu.edu.cn}  
\author[orcid=0000-0003-3901-8403]{Wei Wang} 
\altaffiliation{wangwei2017@whu.edu.cn}
\affiliation{Department of Astronomy, School of Physics and Technology, Wuhan University, Wuhan 430072, China}
\email{wangwei2017@whu.edu.cn}
\author[orcid=0009-0002-6230-8791]{Qianhan Zhou}
\affiliation{Department of Physics, Applied Physics, and Astronomy, Rensselaer Polytechnic Institute, Troy, NY 12180, USA}
\email{zhouq4@rpi.edu}

\begin{abstract}
Based on Insight-HXMT observations, we present a detailed timing analysis and spectral evolution of a complete Main High state for Her X-1 in February 2020. We determine an accurate local ephemeris using the Rømer delay measured from five eclipses. We report the spin period of the neutron star at $P_{\rm spin}=1.23765212 \pm 0.00000026$ s with a spin period derivative of $\dot P_{\rm spin}=-(1.18\pm 0.04)\times 10^{-13}$ s\,s$^{-1}$. By combining the newly measured local values $T_{ecl}$ with those reported in the literature, we refine the orbital ephemeris of Her X-1, obtaining $T_{ecl} = 46359.871956 \pm 0.000010$ MJD and $P_{orb}=1.7001674990 \pm 0.0000000105$ day, then detect a continuous decrease in the orbital period with a rate of $\dot{P}_{\rm orb} = -(1.957 \pm 0.335)\times10^{-11}\,\mathrm{d\,d^{-1}}$. We also investigate the evolution of X-ray spectral parameters during the Main High state. The hydrogen absorption column density $N_{\rm H}$ increased monotonously during the phase, and the photon index kept nearly constant. The cyclotron absorption line was detected with a centroid energy around 38 keV, showing no significant evolution with luminosity. The spectral variations with the superorbital phase are discussed within the accretion disk precession scenario.
\end{abstract}
\section{Introduction} \label{sec:intro}
Hercules X-1 (Her X-1), an eclipsing binary X-ray pulsar with a spin period of 1.24 s in a circular orbit of $\sim 1.7$ d, was first discovered in 1972 during Uhuru observations \citep{giacconi1973further}. Shortly after its discovery, the blue variable 13th-magnitude star HZ Her was identified as its optical companion \citep{davidsen1972identification} and the system's distance has been estimated to be $6.6 \pm 0.4$ kpc \citep{reynolds1997new}. The Her X-1/HZ Her binary system exhibits a 35-day superorbital cycle showing two maxima in intensity: the main-high (MH) state and the short high (SH) state, with the MH lasting approximately 10 days and the SH with flux reaching about 30\% of the MH maximum lasting about 5 days, and separated by low states (LS) of roughly 10 days with flux at approximately 1\% of the maximum (\citealt{giacconi1973further},\citealt{scott1999rossi},\citealt{igna2011hercules}).
\par
The X-ray emission from the Her X-1/HZ Her system arises from several components \citep{leahy2022spectral}. The primary contribution comes from direct X-ray beams emitted by the accretion column of Her X-1 \citep{leahy2004mass}. In addition, weak isotropic emission produced by Thomson scattering of direct radiation occurs at higher altitudes above the column \citep{scott200035}. A soft component below 1 keV, which can be modeled by a blackbody with a temperature of $kT\approx 0.1\ keV$ results from the reprocessing of hard X-rays in the outer magnetosphere of the neutron star and the accretion disk \citep{mccray1982einstein}. The spectrum also includes reflected emission from the inner edge of the disk \citep{leahy2002modelling} and the irradiated face of the companion star \citep{leahy1999extreme}. Finally, unpulsed scattered component is produced by the extended accretion-disk corona, which is visible throughout the 35-day cycle \citep{leahy2015hercules}. Due to the precession of the nearly edge-on accretion disk, these emission regions are periodically obscured and revealed every 35 days, causing the central neutron star's X-ray emission to be blocked and exposed in the 35-day cycle (\citealt{petterson1975hercules},\citealt{klochkov2006observational},\citealt{leahy2002modelling}). \citep{deeter199835} revealed that an occultation effect caused by a precessing inner accretion disk leads to the systematic evolution of the pulse profile in Her X-1 over a 35-day cycle. The correlation between the Main-High state and the neutron star spin period indicates an evolution in angular momentum transfer from the disk \citep{staubert2009two}. Other manifestations consistent with the 35-day cycle include the white-noise process in the first derivative of the 35-d phase fluctuations \citep{baykal1993noise}, and the variations in the optical light curve of HZ Her \citep{kolesnikov2020modelling}.
\par
The overall spectral shape of Her X-1, similar to that of other accreting X-ray pulsars, can be described by a power-law continuum with an approximately exponential cut-off above 10 keV \citep{ji2019long,inam2005x}, a broad Fe K$\alpha$ emission feature at $\sim$ 6.5 keV \citep{choi1994iron} and a prominent cyclotron line appears around 40 keV, first discovered in 1977 through balloon observations and now known as the cyclotron resonant scattering feature (CRSF) \citep{trumper1978evidence}. Long-term monitoring indicates that the CRSF energy gradually decreased since 2006–2009, reaching approximately 37 keV, and remained stable from 2012 to 2020, which may reflect a reconfiguration of the magnetic field and a change in the geometry of the accretion column \citep{xiao2023insight,fuerst2013smooth}. The remarkable 35-day superorbital cycle of Her X-1 is accompanied by changes in its complex spectral characteristics \citep{leahy2022spectral}. Based on RXTE/PCA observations of Her X-1 during the MH state, \citet{leahy2022spectral} reported systematic variations of several spectral parameters over the 35-day cycle. The photon index decreases from values above 1.0 to approximately 0.85, while the cutoff energy declines from $\sim$20.5 keV to $\sim$18 keV between 35-day phases $\sim$0.05 and 0.22 (the zero phase of the 35-day cycle is defined at the turn-on of the high state). In addition, the Fe K$\alpha$ line energy exhibits a reduction, indicating a decrease in the ionization state of the fluorescing gas. These observed trends, particularly the delayed iron line variability and the hardening of the spectrum, are broadly consistent with expectations from a precessing accretion disk, in which changes in the viewing geometry modulate the relative contributions of direct and reflected emission from the inner disk edge.
\par
This paper is organized as follows. In Section 2, the observations of Insight-HXMT on Her X-1 are introduced, and the analysis processes of the raw data are also presented. In Section 3, the science data analysis and results are presented, including the pulse period search, pulse profiles and orbital elements. Spectral analysis is presented in Section 4. In Section 5, we discuss the long-term evolution of the orbital period and the spectral properties of Her X-1 during the Main High State. The conclusion is
presented in Section 6.

\section{OBSERVATIONS}
\label{OBSERVATIONS}
\begin{figure*}
    \centering
    \includegraphics[width=0.75\textwidth, height=0.5\textwidth]{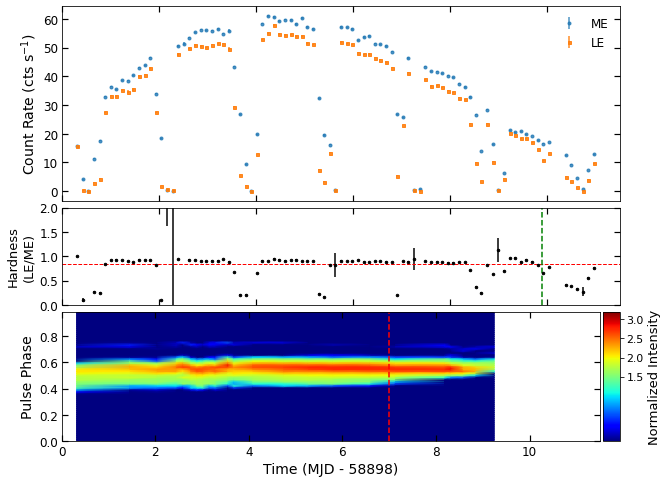}
    \caption{The top panel shows the background-subtracted X-ray light curves of Hercules X-1 observed by Insight-HXMT between 19 and 29 February 2020, with a time resolution of 10,000 s. Orange points represent the LE (1–10 keV) countrate, and blue points represent the ME (10–20 keV) countrate. The middle panel shows the softness ratio (SR), defined as the LE countrate divided by the ME countrate, with its mean value indicated by a red dashed line. The green dashed line marks the onset of spectral hardening around MJD 58908. The bottom panel shows the evolution of the ME (10–20 keV) pulse profiles during the Main High State. After MJD 58905, the shoulder peak in the pulse phase range 0.4–0.45 disappears, whereas the central peak is still present.}
    \label{fig:1}
\end{figure*}

The Hard X-ray Modulation Telescope (Insight-HXMT) is the first Chinese X-ray astronomical satellite launched on 2017 June 15. Insight-HXMT consists of three main instruments: the High Energy X-ray telescope (HE) operating in 20--250 keV and the geometrical areas of the telescopes are 5100 cm$^2$ \citep{liu2020high}, the Medium Energy X-ray telescope (ME) operating in 5--30 keV with a geometrical detection area of 952 cm$^{2}$ \citep{cao2020medium} and the Low Energy X-ray telescope (LE) covering the energy range 1--15 keV with a geometrical detection area of 384 cm$^2$ \citep{chen2020low}. Since its launch, Insight-HXMT has monitored Her X-1 through multiple observations between July 2017 and January 2025. The dataset used in this work corresponds to a single complete MH state (cycle 505; \citealt{xiao2023insight}). The corresponding light curve is shown in Figure \ref{fig:1}. The Insight-HXMT Data Analysis Software (HXMTDAS) v2.04 is used to analyze data (more details on the analysis were introduced in previous publications, e.g., \citealt{WANG20211}; \citealt{yang2024evidence}). To take advantage of the best-screened event file to generate the high-level products including the energy spectra, response file, light curves and background files, we use tasks $he/me/lepical$ to remove spike events caused by electronic systems and $he/me/legtigen$ be utilized to select good time interval (GTI) when the pointing offset angle $< 0.04^\circ$; the pointing direction above earth $> 10^\circ$; the geomagnetic cut-off rigidity $>8$ GeV and the South Atlantic Anomaly (SAA) did not occur within 300 seconds. Tasks helcgen, melcgen, and lelcgen are used to extract X-ray lightcurves with $128^{-1}$ sec time bins. We use the XSPEC v12.12.0 software package \citep{arnaud1996astronomical} included in HEASoft v6.29 for spectral fitting and error estimation. 
\par
\section{ANALYSIS AND RESULTS}
\subsection{Light-curve and pulse profiles}
Figure \ref{fig:1} shows the background-subtracted light curves of Her X-1 with a time resolution of 10000 s, observed by Insight-HXMT between 19 and 29 February 2020. The orange points represent data in the LE (1-10 keV), while the blue points correspond to the ME (10-20 keV). We calculated the Softness Ratio (SR), defined as the LE count rate divided by the ME count rate, and indicated its mean value with a red dashed line. Seven eclipses are clearly visible during this interval, with the source becoming fainter and the spectrum hardening during each eclipse. This behavior is similar to that reported by \citet{abdallah2015spectral} during the LS, where a significant hard spectral component modulated over the binary orbit was attributed to the reflection from the irradiated face of the companion star HZ Her. The spectrum hardening observed during eclipses in our Insight–HXMT data indicates that the reflection signal is still apparent even when the direct emission is strong. The SR also shows that after $\sim$MJD 58908, the source exhibited spectrum hardening. However, owing to the absence of low-state observations, the detailed evolution of this behavior cannot be well constrained. As the 35-day phase progresses toward the end of the MH, the observed spectrum hardening could be attributed to an increasing contribution from reflected emission from the inner region of the accretion disk relative to the direct emission from the central neutron star. As the reflected spectrum is expected to be harder (flatter) at high energies ($>$10 keV), this may explain the apparent hardening of the phase-averaged spectrum \citep{leahy2022spectral}. 
\par
The corresponding pulse profiles in 10–20 keV, folded with the period of $\sim$1.24 s which is determined by using the $efsearch$ method with 1/128 s, are also shown in the bottom panel of Figure \ref{fig:1}. The X-ray pulse profile undergoes a systematic evolution during the Main High State. The profile is composed of a central peak (phase range 0.45–0.5), a shoulder peak (phase range 0.4–0.45) and trailing shoulders around phase 0.7. After MJD 58905, the shoulder peak disappears, whereas the main peak is still present. This evolution is best explained by the disk sequentially obscuring different parts of the pulsar beam, causing the characteristic variation of the pulse components \citep{scott200035}.

\subsection{Spin evolution and orbital parameters}
Using the $efsearch$ method, the observed spin period ($P_{obs}$), which is affected by the Doppler shift due to the binary motion is obtained from each observation. The uncertainties of $P_{obs}$ are estimated using a Gaussian error. To derive the intrinsic spin period ($P_{spin}$) from the observed values, the Doppler effects from the binary orbit must be corrected \citep{fu2023timing,weng2017swift}. Consequently, we use the pulse period evolution to infer the orbital parameters \citep{susobhanan2018exploring}:
\begin{equation}
P_{obs}\left( t \right) \approx [P_{\rm spin}(t_0) + (t - t_0) \, \dot{P}_{\rm spin}]\left( 1+\frac{d\bigtriangleup_R}{dt} \right),
\end{equation}
where $\dot{P}_{\rm spin}$ is the derivative of the spin period, \textbf{and} $t_0$ is a reference time. The Rømer delay $\bigtriangleup_R$, associated with an eccentric Keplerian orbit, is given by \citep{blandford1976arrival}
\begin{equation}
\bigtriangleup _R=\frac{a_x\sin i}{c}\left[ \left( \cos \nu -e \right) \sin \omega +\sqrt{1-e^2}\sin \nu \cos \omega \right].
\end{equation}
Here, $a_x \sin i$ represents the semimajor axis of the pulsar's orbit projected along the line of sight, where $i$ denotes the orbital inclination. The Keplerian orbital parameters include the orbital eccentricity $e$, the true anomaly $\nu$, and the argument of periastron $\omega$.
Their relationships are expressed as follows:
\begin{equation}
\nu = 2 \arctan\left( \frac{\sqrt{1+e} \, \sin\frac{E}{2}}{\sqrt{1-e} \, \cos\frac{E}{2}} \right),
\end{equation}
\begin{equation}
E-e\sin E=\frac{2\pi}{P_{orb}}\left( t-T_{ecl} \right).
\end{equation}
$E$ in Equation (4) denotes the eccentric anomaly, while $T_{\rm ecl}$ denotes the time at which the neutron star's mean orbital longitude is $90^\circ$, corresponding to the maximum pulse arrival time delay or a zero pulse period derivative. In Her X-1, this is essentially identical to the center of the eclipse \citep{staubert2009updating}. When the orbit is nearly circular, with the eccentricity reported as $e \approx 0.00042$ \citep{staubert2009updating}, the orbital equations can be approximated as \citep{liu2023measurements}:
\begin{equation}
\begin{aligned}
P_{\rm obs}(t) &= \big[P_{\rm spin}(t_0) + (t - t_0) \dot{P}_{\rm spin}\big] \\
&\quad \times \Big[ 1 + \frac{2 \pi a_x \sin i}{P_{\rm orb}} 
\sin \Big( \frac{2 \pi (t - T_{\rm ecl})}{P_{\rm orb}} \Big) \Big].
\end{aligned}
\end{equation}
\begin{figure*}
    \centering
    \includegraphics[width=0.8\textwidth]{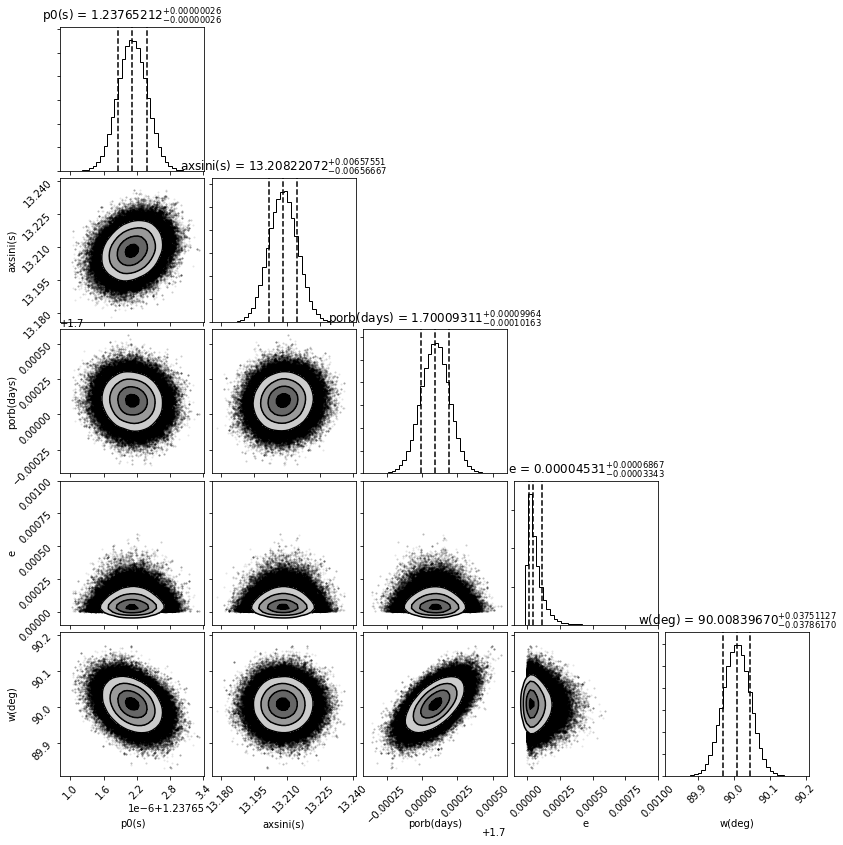}
    \caption{Corner plot of the MCMC posterior distributions of the orbital parameters of Her X-1 assuming an eccentric orbital model. The diagonal panels show the one-dimensional distributions with the median and $1\sigma$ intervals, while the off-diagonal panels show the joint distributions. The small value of $e$ indicates that the orbit is nearly circular.}
    \label{fig:15}
\end{figure*}
In this work, we first fit the pulse periods assuming a circular orbit using the above approximation, and then consider the full eccentric Keplerian model for completeness. To ensure a sufficient number of data points and to reduce the averaging effect on the spin period caused by a long time, we divided each observation into multiple segments. Each segment was chosen to have a duration of at least 500 seconds, and the midpoint of the segment was taken as the time for the measured $P_{obs}$. We use a Bayesian approach with Markov Chain Monte Carlo (MCMC) to sample the posterior distribution of the model parameters, obtaining their most likely values and 1$\sigma$ confidence intervals.  When we considered the eccentric orbit, we found that $T_{\rm ecl}$ and $\omega$ are strongly positively correlated because both parameters influence the timing of the pulse arrivals in a very similar way, so an increase in $\omega$ can be largely offset by a corresponding shift in $T_{\rm ecl}$, making it difficult to independently constrain them.  To address this degeneracy, $T_{\rm ecl}$ was fixed to the value obtained from the circular orbit fit, while the other parameters of the eccentric orbit were left free to vary. Figure~\ref{fig:15} shows the corner plot of the best-fitted results. Most parameters exhibit nearly Gaussian one-dimensional posterior distributions and are well constrained. In particular, the projected semi-major axis $a_x\sin i$, the orbital period $P_{\rm orb}$, and the spin period $P_{\rm spin}$ are tightly constrained. Notably, the posterior distribution of the eccentricity $e$ is concentrated within a very small positive range and shows a slightly asymmetric shape. The MCMC sampling yields $e = (4.53^{+6.87}_{-3.34}) \times 10^{-5}$,
indicating that the orbital eccentricity is resolved but extremely small. Its $1\sigma$ upper limit is approximately $e \lesssim 1.1 \times 10^{-4}$, suggesting that the system is in a highly circularized orbit with only a very small residual eccentricity. 

\begin{figure}
    \centering
    \includegraphics[width=.48\textwidth]{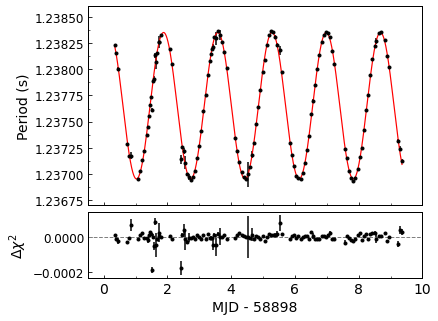}
    \caption{The spin period of Her X-1 observed by Insight-HXMT. The red solid curve represents the best-fitting model. The residuals are shown in the lower panel.}
    \label{fig:2}
\end{figure}

We did not include the spin-period derivative $\dot P_{\rm spin}$ as a free parameter in the MCMC fitting. Instead, we first obtained the intrinsic spin period $P_{\rm spin}$ for each observation by fitting the orbital model. The long-term evolution of $P_{\rm spin}$ was then analyzed by performing a linear regression with time to independently estimate $\dot P_{\rm spin}$. The reference epoch for $P_{\rm spin}$ is adopted from \citet{staubert2009updating}, where $P_{\rm spin} = 1.237739612(25)$ at MJD 50290.659199 \citep{staubert2009updating}. This two-step procedure helps to avoid parameter degeneracy between the long-term spin evolution and the orbital modulation, leading to a more robust orbital solution. To illustrate the fitting performance of the model to the observational data more clearly, we further present a comparison between the best-fitting orbital model and the spin period evolution with time, as shown in Figure~\ref{fig:2}. The best-fitting parameters are listed in Table~\ref{tab:2.2}.

\begin{figure}
    \centering
    \includegraphics[width=0.45\textwidth]{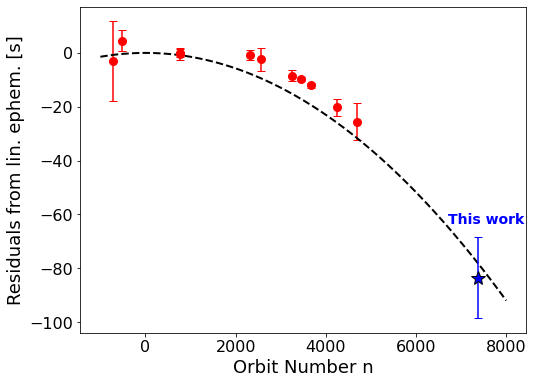}
    \caption{Residuals of $T_{ecl}$ after subtracting the linear component of the best-fitting quadratic ephemeris. The blue pentagrams indicate the results from this work, while the black dashed line represents the best-fitting quadratic ephemeris.}
    \label{fig:17}
\end{figure}

\begin{table}
\centering
\caption{MCMC fitting results of the eccentric orbital model (errors correspond to the $1\sigma$ confidence intervals).}
\label{tab:2.2}
\begin{threeparttable}
\begin{tabular}{lcc}
\hline\hline
Parameter & Unit & Value \\
\hline
$P_{\rm spin}$ & s 
& $1.23765212 \pm 0.00000026$ \\
$\dot P_{\rm spin}$ & s\,s$^{-1}$ 
& $-(1.18\pm0.04) \times 10^{-13}$ \\
$a_x \sin i$ & lt-s 
& $13.20822 \pm 0.00657$ \\

$P_{\rm orb}$ & day 
& $1.70009311 \pm 0.00010$ \\

$e$ & -- 
& $(4.53^{+6.87}_{-3.34}) \times 10^{-5}$ \\

$\omega$ & deg 
& $90.0084 \pm 0.0377$ \\

$T_{ecl}$\tnote{a} & MJD 
& $58902.006$(fixed) \\
\hline
\end{tabular}
\begin{tablenotes}
\footnotesize
\item[a] $T_{ecl}$ was first obtained by fitting the circular orbital model ($e=0$). It was then kept fixed during the MCMC sampling of the eccentric orbital model in order to reduce the parameter degeneracy between $T_{ecl}$ and $\omega$ and to improve the stability of the orbital solution.
\end{tablenotes}
\end{threeparttable}
\end{table}
Based on our updated $T_{ecl}$ measurements (Table~\ref{tab:2.2}) together with archival observations spanning the past $\sim$40 years \citep{deeter1991decrease, staubert2009updating}, we re-estimate the orbital period derivative $\dot{P}_{\rm orb}$ of Her X-1. Under a quadratic ephemeris model, the eclipse epoch of the $n$-th orbit
at an arbitrary observation time satisfies \citep{staubert2009updating}:
\begin{equation}
    T_{ecl}(n) = T_{ecl}(0) + n\,P_{\rm orb}(0) + 
    \frac{1}{2} n^2\,P_{\rm orb}(0)\,\dot{P}_{\rm orb}(0),
\end{equation}
where $T_{ecl}(0)$ corresponds to the reference epoch when $n=0$,
and $P_{\rm orb}(0)$ is the orbital period at the reference time
$T_{ecl}(0)$. The best-fitting parameters of the quadratic ephemeris are listed in Table~\ref{tab:2.3}. After subtracting the linear component of the best-fitting quadratic ephemeris, the residuals of $T_{ecl}$ are shown in Figure~\ref{fig:17}.
\begin{table}[h!]
\centering
\caption{Best-fitting parameters of the quadratic ephemeris for Her X-1  (errors correspond to the $1\sigma$ confidence intervals)}
\label{tab:2.3}
\begin{tabular}{lc}
\hline\hline
Parameter & Value \\
\hline
$T_{ecl}$ [MJD] & $46359.871956 \pm 0.000010$ \\
$P_{\rm orb}$ [d] & $1.7001674990 \pm 0.0000000105$ \\
$\dot{P}_{\rm orb}$ [d/d] & $-(1.957\pm 0.335)\times10^{-11} $ \\
\hline
\end{tabular}
\end{table}
\section{Spectral Analysis}
Common empirical models can generally describe the hard X-ray spectra of accreting pulsars reasonably well. However, studies of Her X-1 using the PCA and HEXTE instruments on board RXTE have shown that when adopting cutoffpl or NPEX as the continuum model, strong correlations exist between the spectral parameters. Such correlations increase the difficulty of constraining parameters during spectral fitting and result in larger uncertainties in the derived spectral parameters \citep{coburn2002magnetic}. Therefore, in the spectral analysis of this work, we adopt the powerlaw × highecut model to describe the continuum part of the
spectra of Her X-1, which is consistent with the choice of continuum models in previous spectral studies of Her X-1 \citep{xiao2024insight, leahy2022spectral, vasco2013pulse}.
Based on the continuum model, we further include a Gaussian emission component to account for the Fe K$\alpha$ fluorescent line. To describe possible absorption effects at low energies, two different modeling approaches can be adopted: a full-covering absorption model (e.g., tbabs) and a partial-covering absorption model (e.g., tbpcf). Systematic studies of Her X-1 have shown that in the $2.5$–$30\,keV$ energy range, the partial-covering absorption model statistically provides a better description of the spectral shape than the full-covering model, yielding smaller $\chi^2$ values in most pulse phases. Therefore, it is considered \textbf{to be} the most reasonable model for describing the X-ray spectrum of Her X-1 and has been widely applied in previous studies \citep{leahy2022spectral}. Based on these results, we adopt the partial-covering absorption model to account for interstellar absorption in the spectral fitting of this work. However, the background of the HE detector is more than five times higher than the source flux above $\sim 60\, keV$\citep{xiao2024insight}. Considering these limitations, we set the upper energy limit of the spectral fitting to $60\,keV$. To improve the statistics at high energies, we combine observations using the addspec tool in XSPEC\footnote{https://heasarc.gsfc.nasa.gov/docs/software/lheasoft/help/addspec.html}
, thereby increasing the photon counts in the high-energy band.
\par
\begin{figure}[t!]
    \centering
    \includegraphics[width=0.5\textwidth, height=10cm]{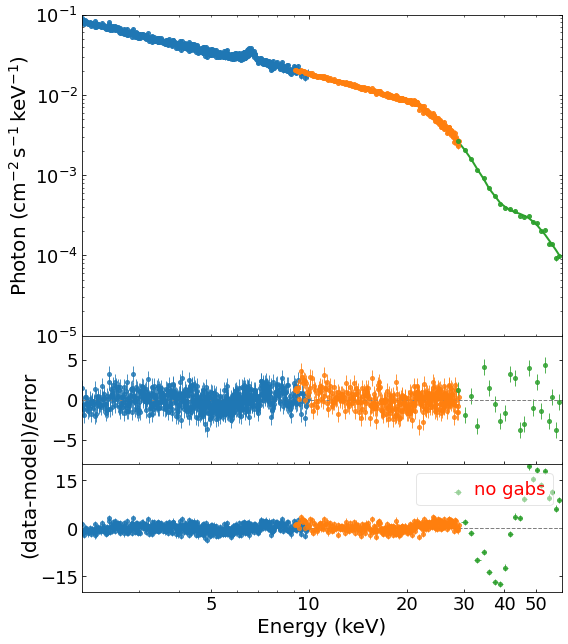}
    \caption{Spectral fitting results of Her X-1 in the 2–60 keV energy range, obtained by combining five spectra at the Main-On peak phase. The middle panel shows the corresponding fitting residuals, and the bottom panel shows the fit without the cyclotron absorption line component.}
    \label{fig:64} 
\end{figure}

To verify previous studies indicating that the cyclotron resonance scattering feature (CRSF) of Her X-1 is centered around $\sim37 \,keV$, we first fitted the spectra without including a gabs component. A clear absorption feature appeared in the 30–40 keV range, with a resulting fit of $\chi^2 = 4797.58$ (1297 d.o.f). Consequently, we added a gabs component to model the CRSF, which reduced $\chi^2$ to 1716.15 (1294 d.o.f). An example spectrum and the fitting procedure are shown in Figure~\ref{fig:64}, with the corresponding parameters listed in Table~\ref{tab:4.6}.
\begin{table}[htbp]
\centering
\caption{Spectral fitting parameters of Her X-1 in the 2–60 keV energy range at the Main-On peak phase (combining five spectra). Errors correspond to the 90$\%$ confidence interval.}
\resizebox{0.5\textwidth}{!}{
\begin{tabular}{lcc}
\hline
Component & Parameter & Value \\
\hline
\texttt{pcfabs} & $N_{\rm H}$ ($10^{22}$ cm$^{-2}$) & $24.43_{-1.95}^{+1.96}$ \\
                & Covering Fraction & $0.299_{-0.026}^{+0.030}$ \\

\texttt{highecut} & $E_{\rm cut}$ (keV) & $21.14_{-0.14}^{+0.20}$ \\
                  & $E_{\rm fold}$ (keV) & $11.71_{-0.17}^{+0.18}$ \\

\texttt{powerlaw} & Photon Index & $1.145_{-0.020}^{+0.023}$ \\
                  & Norm & $0.272_{-0.015}^{+0.018}$ \\

\texttt{gaussian} & Line Energy (keV) & $6.627_{-0.025}^{+0.028}$ \\
                  & $\sigma$ (keV) & $0.167_{-0.038}^{+0.066}$ \\
                  & Norm & $(2.78_{-0.43}^{+0.48})\times10^{-3}$ \\

\texttt{gabs} & Line Energy (keV) & $38.50_{-0.22}^{+0.28}$ \\
              & $\sigma$ (keV) & $5.31_{-0.27}^{+0.37}$ \\
              & Strength & $9.34_{-0.64}^{+0.85}$ \\

\hline
$L_{\rm X}$ (2--60 keV) & $10^{37}\,erg\,s^{-1}$ & $3.147_{-0.017}^{+0.018}$ \\
\hline
\label{tab:4.6}
\end{tabular}}
\end{table}
\par
At energies below 30 keV, the continuum parameters can generally be constrained relatively well, whereas at energies above 30 keV, the lower signal-to-noise ratio may affect the determination of these parameters \citep{leahy2022spectral}. We used Insight-HXMT LE and ME data to fit the phase-averaged spectra for each observation to constrain the continuum parameters, with the fitting energy ranges set to 2–10 keV for LE and 10–30 keV for ME. Within the 2–30 keV energy range, we ultimately adopted the $(\mathrm{tbpcf} \times \mathrm{powerlaw} \times \mathrm{highecut}) + \mathrm{gaussian}$ model to fit the phase-averaged spectra of Her X-1.
\par
Since the $E_{\rm fold}$ describes the exponential decay scale of the continuum spectrum above the cutoff energy ($E_{\rm cut}$), \textbf{it characterizes} the decay rate of the high-energy spectral tail. When only fitting spectra below 30 keV, the high-energy exponential decay is not sufficiently sampled, which leads to a certain degeneracy between $E_{\rm fold}$ and other continuum parameters. To reproduce the observed count rate decline within a limited energy band, the fitting process tends to a smaller $E_{\rm fold}$ value to compensate for the lack of high-energy data \citep{leahy2022spectral}. Therefore, in order to reduce parameter degeneracy and obtain more stable continuum fits, we fixed $E_{\rm fold}$ at 10 keV in subsequent fits, which lies within the typical range of $E_{\rm fold}$ values reported in previous studies of Her X-1 spectra \citep{xiao2024insight}.

\par
\begin{figure*}[htbp]
\centering

\begin{subfigure}{0.48\textwidth}
    \includegraphics[width=\linewidth]{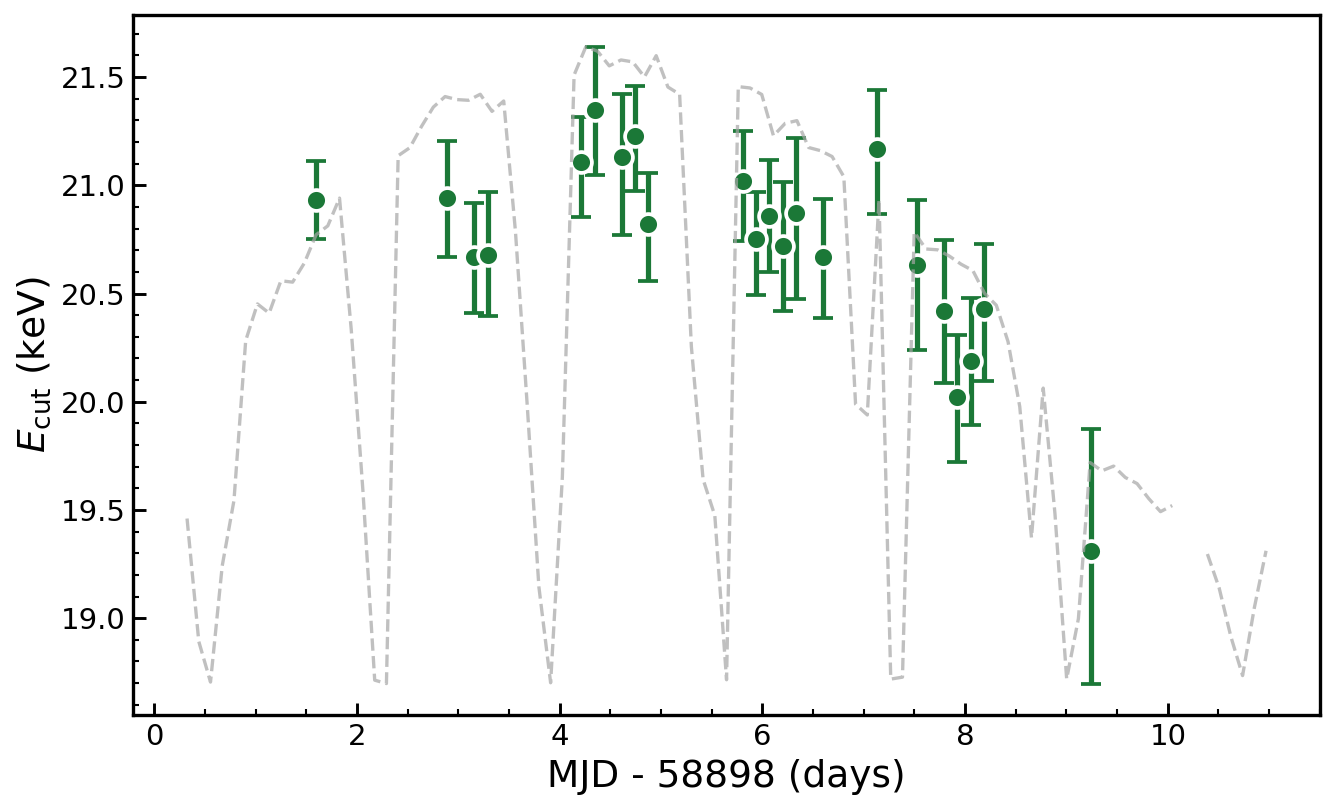}
    \caption{}
\end{subfigure}
\hfill
\begin{subfigure}{0.48\textwidth}
    \includegraphics[width=\linewidth]{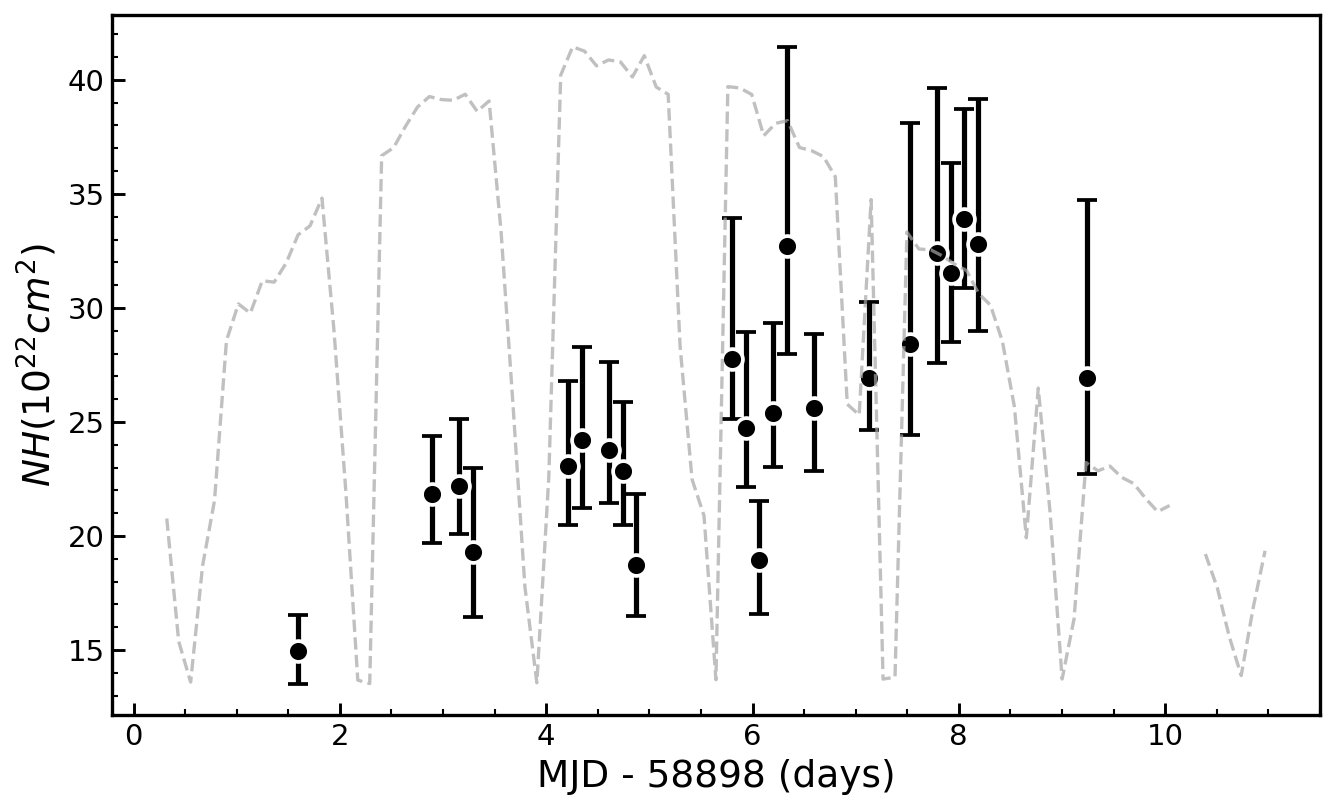}
    \caption{}
\end{subfigure}

\vspace{0.3cm}

\begin{subfigure}{0.48\textwidth}
    \includegraphics[width=\linewidth]{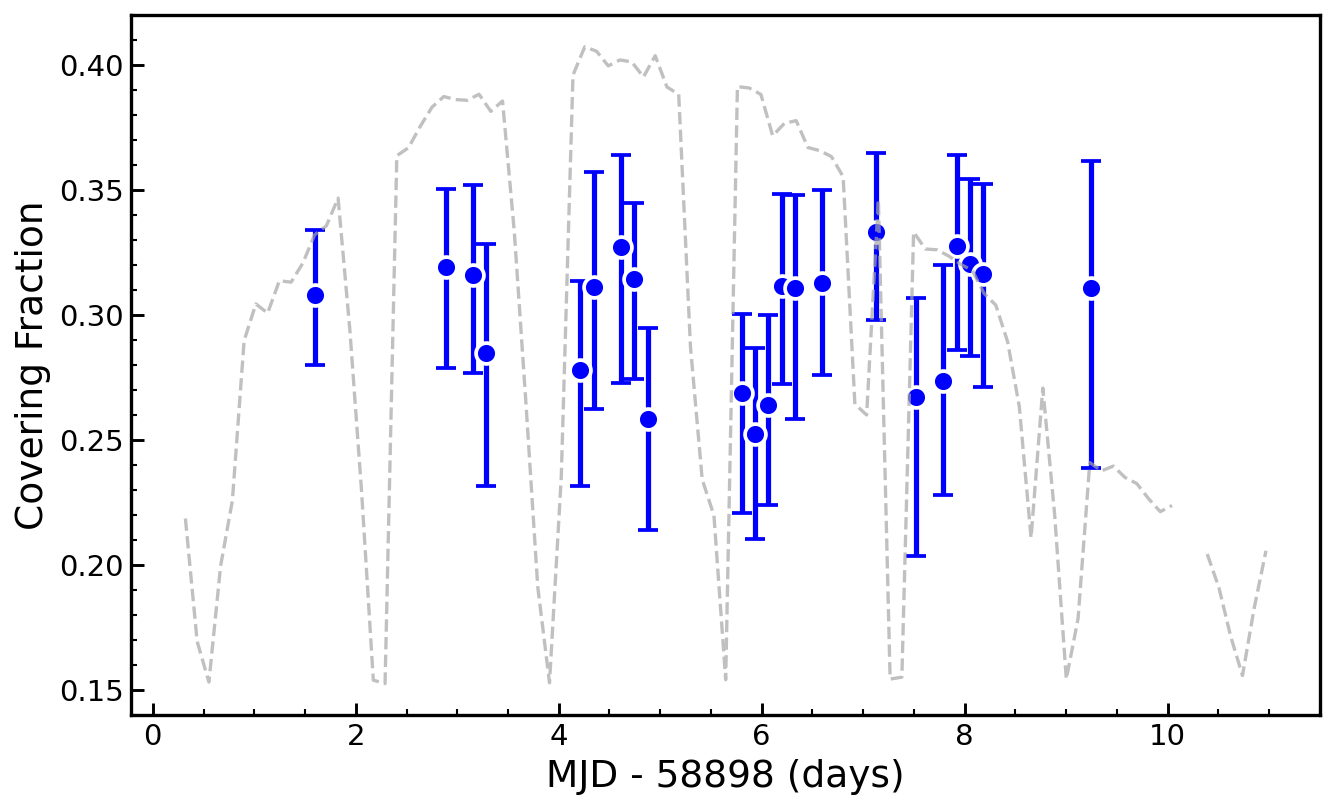}
    \caption{}
\end{subfigure}
\hfill
\begin{subfigure}{0.48\textwidth}
    \includegraphics[width=\linewidth]{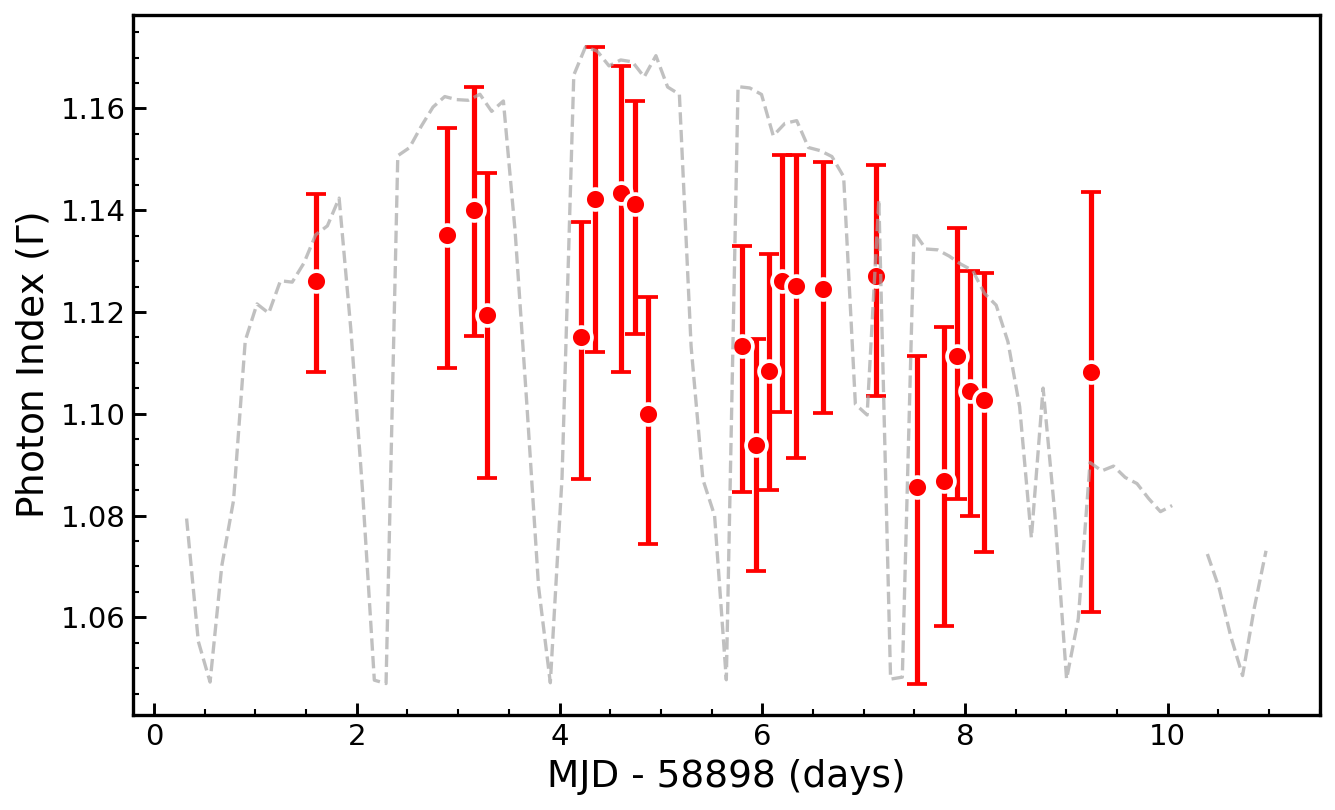}
    \caption{}
\end{subfigure}

\vspace{0.3cm}

\begin{subfigure}{0.48\textwidth}
    \includegraphics[width=\linewidth]{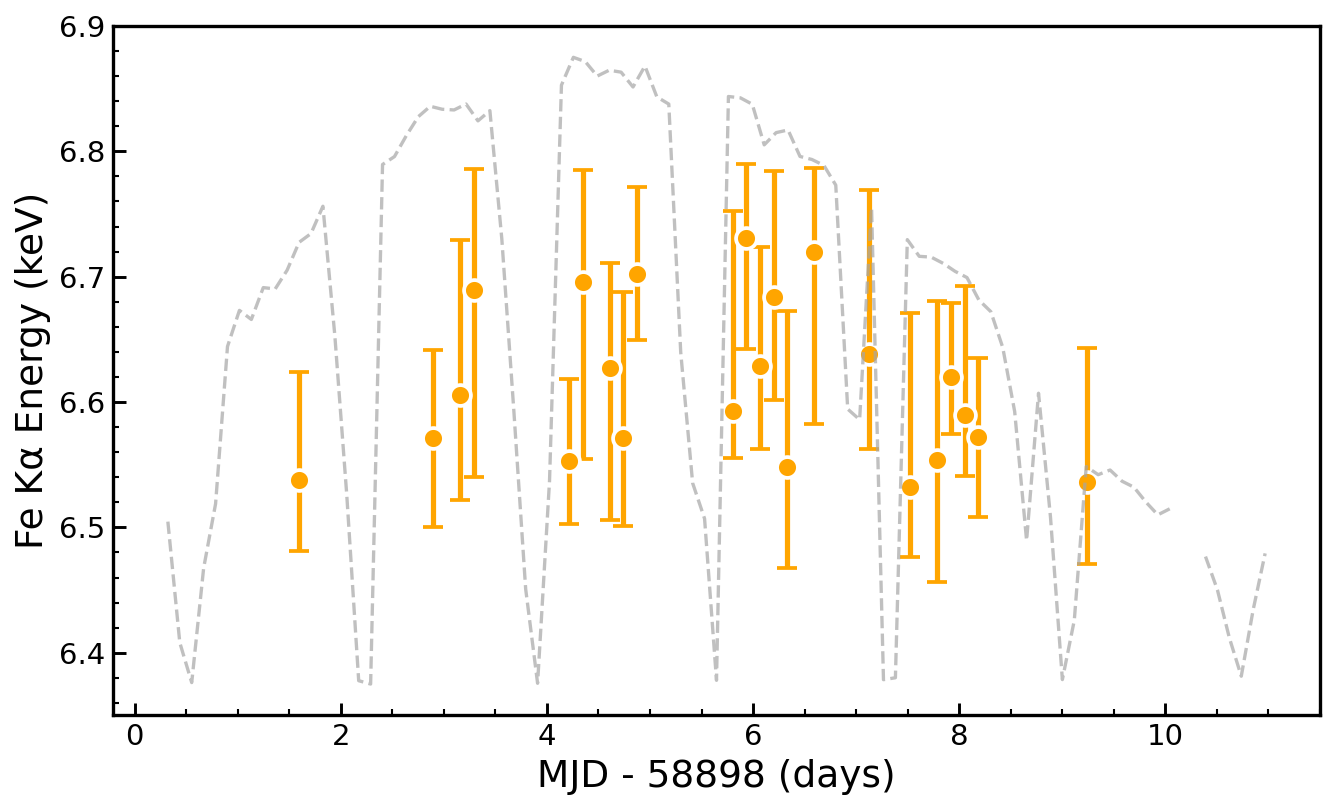}
    \caption{}
\end{subfigure}
\hfill
\begin{subfigure}{0.48\textwidth}
    \includegraphics[width=\linewidth]{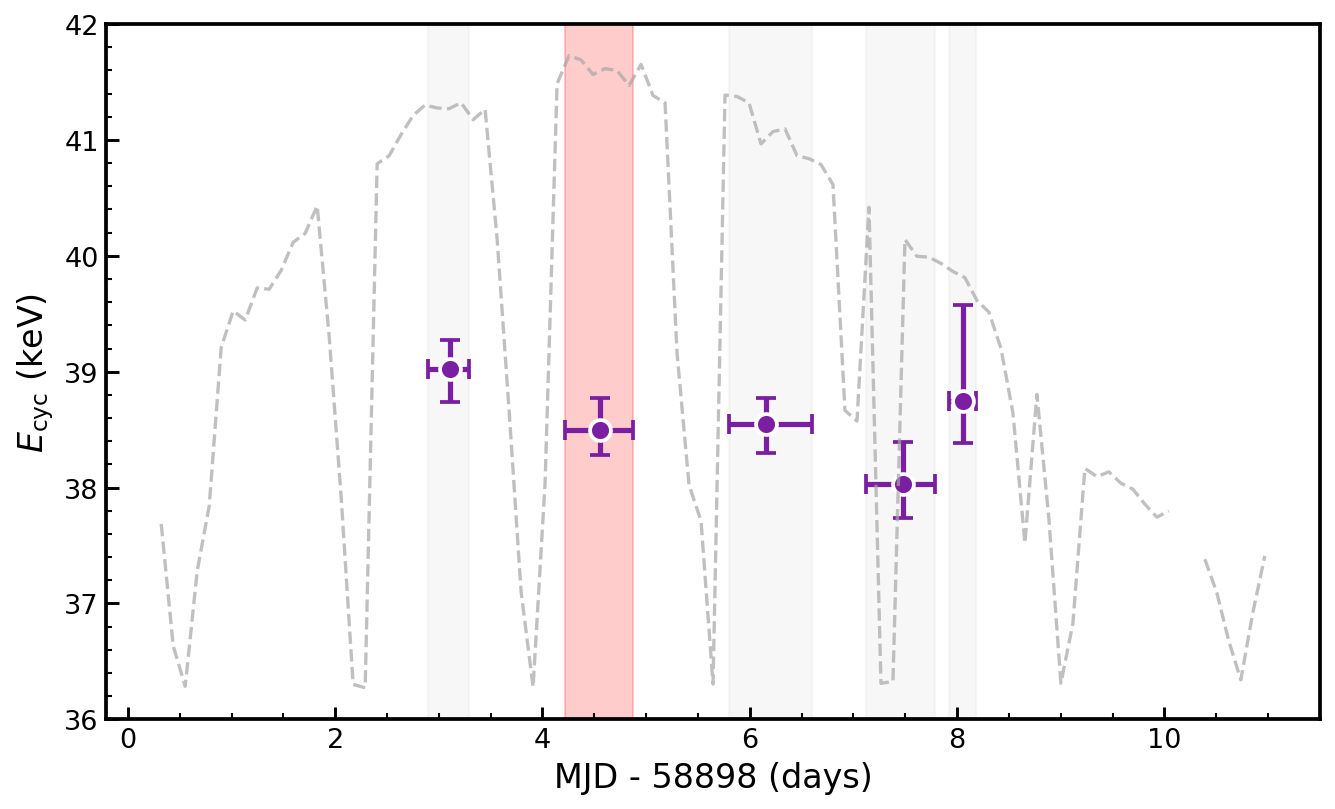}
    \caption{}
\end{subfigure}

\caption{Evolution of Her X-1’s spectral parameters over the Main-High phase. The panels show the spectral parameters $E_{\rm cut}$ (a), $N_{\rm H}$ (b), covering fraction (c), power-law index (d), and Fe K$\alpha$ centroid energy (e). Panel (f) shows the cyclotron line centroid energy. The combined time intervals used in Panel (f) are indicated in the figure by gray and red shaded regions, where the red shaded region corresponds to the spectral fit shown in Figure \ref{fig:64}, with the relevant fit parameters listed in Table \ref{tab:4.6}.}
\label{fig:multi_panel}
\end{figure*}
\par
We investigated the temporal evolution of the spectral parameters of Her X-1 over the Main-High phase, and the results are shown in Figure~\ref{fig:multi_panel}. Overall, the spectral parameters exhibit certain evolutionary trends during this stage. The equivalent hydrogen column density \(N_\mathrm{H}\) generally increases during the Main High state (Figure~\ref{fig:multi_panel}b), rising from roughly \(15 \times 10^{22}~\mathrm{cm}^{-2}\) to approximately \(35 \times 10^{22}~\mathrm{cm}^{-2}\). The covering fraction fluctuates between about 0.25 and 0.35 (Figure~\ref{fig:multi_panel}c). The photon index \(\Gamma\) varies slightly over the entire observation period, remaining roughly in the range 1.07--1.14 (Figure~\ref{fig:multi_panel}d), indicating that the overall shape of the continuum spectrum remained largely stable between MJD 58899 and 58908. At the same time, the cutoff energy \(E_\mathrm{cut}\) shows a constant value around \(21~\mathrm{keV}\). and in the end of MH phase, decreases to about \(19.5~\mathrm{keV}\). Similarly, the hardness ratio analysis reveals that the source exhibits spectral hardening starting around MJD 58908 (Figure~\ref{fig:1}). It is worth noting that the photon count rate decreased significantly during this period, resulting in a lower signal-to-noise ratio in the spectra and preventing reliable spectral fitting and further analysis. Moreover, due to the lack of low-state observations, the detailed evolution of this spectral hardening cannot be determined. The energy of the Fe K\(\alpha\) emission line is concentrated around 6.5--6.7~keV (Figure~\ref{fig:multi_panel}e). In addition, the CRSF energy \(E_\mathrm{cyc}\) remains around \(\sim 38~\mathrm{keV}\) near the peak of the Main High state. 
\section{Discussion}
The Insight-HXMT observations have extended the time baseline for determining the orbital epoch of Her X-1. Consequently, we report a new value for the orbital period derivative: 
\(\dot{P}_{\rm orb} = (-1.957 \pm 0.335) \times 10^{-11}~\rm{d/d}\). 
Compared with previous studies, \(\dot{P}_{\rm orb} = -(4.85 \pm 0.13) \times 10^{-11}~\rm{d/d}\) \citep{staubert2009updating} and 
\(\dot{P}_{\rm orb} = (-6.16 \pm 0.74) \times 10^{-11}~\rm d/d\) \citep{deeter1991decrease}, the absolute value of our updated orbital period derivative is slightly lower than earlier reports, but still within the theoretically allowed range, indicating that this change is physically reasonable. The observed magnitude of the orbital period change can be explained by the removal of system angular momentum via stellar coronal winds \citep{stelzer1997evolution}.
\par
Based on a simplified conservative mass-transfer model, assuming the neutron star and the optical companion as point masses and considering only orbital angular momentum, the neutron star accretion rate \(\dot{M}_{\rm NS}\) can be related to the orbital period derivative \(\dot{P}_{\rm orb}\) through
\begin{equation}
\frac{\dot{M}_{\rm NS}}{M_{\rm NS}} - \frac{\dot{M}_{\rm NS}}{M_{\rm opt}} + \frac{\dot{P}_{\rm orb}}{3 P_{\rm orb}} = 0,
\end{equation}
where \(M_{\rm NS}\) and \(M_{\rm opt}\) are the masses of the neutron star and the optical companion, respectively, 
and \(\dot{M}_{\rm NS}\) is the neutron star accretion rate. Using \(M_{\rm NS} = 1.4~M_\odot\) and \(M_{\rm opt} = 2.2~M_\odot\), together with our measured \(\dot{P}_{\rm orb}\), we obtain a neutron star accretion rate of \(\dot{M}_{\rm NS} \approx 5.39 \times 10^{-9}\,M_\odot\,{\rm yr}^{-1} \approx 3.4 \times 10^{17}~{\rm g/s}\). According to the classical Newtonian estimate of the neutron star accretion luminosity, the corresponding accretion luminosity is \(L_{\rm acc} \sim 6.3 \times 10^{37}~{\rm erg/s}\), which is still significantly higher than the observed X-ray luminosity \(L_X \sim 3.15 \times 10^{37}~{\rm erg~s^{-1}}\).
\par
To reconcile the observed luminosity, non-conservative mass transfer must be considered, i.e., part of the matter is lost from the system via the companion's wind guided by the magnetic field, carrying away angular momentum. Specifically, we estimate that approximately $50\%$ of the mass flowing from the companion is carried away by the magnetized stellar wind, removing angular momentum from the system. Nevertheless, the conservative mass-transfer model is highly simplified, neglecting the angular momentum of the accretion disk and accretion flow, their dynamical effects on the stars, and the movement of the system’s center of mass during mass transfer, all of which can significantly affect the final result \citep{deeter1991decrease}.
\par
By fitting the phase-averaged spectra, we derived both the continuum spectral parameters and the cyclotron resonance scattering feature  parameters, and investigated their evolution over the 35-day superorbital period. The results show that during the Main High state, the continuum parameters exhibit relatively modest variations but still display a certain evolutionary trend. In particular, the equivalent hydrogen column density \(N_\mathrm{H}\) shows an overall increasing trend during the Main High state, the partial covering fraction varies between \(\sim 0.25\) and \(\sim 0.35\). The photon index \(\Gamma\) remains relatively stable in the range 1.07–1.14, the high-energy cutoff \(E_\mathrm{cut}\) decreases from \(\sim 21\,\mathrm{keV}\) to \(\sim 19.5\,\mathrm{keV}\)  near the end of the MH state.
\par
According to the pulse evolution model proposed by \citet{scott200035}, the observed spectra of Her X-1 during the MH state can be interpreted as a combination of: (1) direct emission from the neutron star (including fan and pencil beams); (2) X-ray emission reflected from the illuminated inner accretion disk ring; and (3) emission scattered in the magnetosphere or corona by the direct radiation. The relative contribution of these components leads to the observed evolution of spectral parameters over the 35-day cycle. The scattered emission from hot plasma may originate in the magnetosphere above the accretion column or in the corona above and below the disk, and its contribution is typically weak and spectrally similar to the direct emission. Thus, during the Main High state, this component contributes little to the overall spectral variability \citep{leahy2022spectral}. In contrast, the reflected emission from the inner accretion disk ring plays a more important role in shaping the observed spectral evolution. As noted by \citet{scott200035}, during the high state the inner disk ring is strongly irradiated by the fan-beam  from the accretion column. In the 35-day pulse evolution model, the reversed fan beam is produced at a height of roughly two neutron star radii above the neutron star surface and subtends a solid angle about 5–10 times larger than that of the pencil beam. As a result, the inner disk ring is primarily 
illuminated by the fan beam, and the reflected emission is therefore 
dominated by radiation originating from the fan beam. As the superorbital cycle progresses, this hot atmosphere gradually reveals the inner disk ring, increasing its observed flux. Toward the end of the MH state, the near side of the inner disk edge again partially blocks the central source and the inner ring. Under this scenario, the evolution of spectral parameters over the 35-day cycle during the MH state can be naturally explained.  
\par
Within this geometrical evolution framework, the changes in absorption structure described by the partial covering model are also reasonably interpreted. As the superorbital phase advances and the inner disk ring becomes more visible, the distribution of material along the line of sight changes, leading to variations in the covering fraction, which fluctuate between \(\sim 0.25\) and \(\sim 0.35\). Meanwhile, the overall increase of \(N_\mathrm{H}\) indicates that the total column density along the line of sight rises as the inner disk ring is revealed.
\par
During the early MH phase, the direct emission dominates, and the phase-averaged spectrum is expected to show a relatively high cutoff energy. As the 35-day cycle progresses, the contribution of reflected and scattered emission increases, causing a gradual decrease in the phase-averaged cutoff energy. \citet{vasco2013pulse} performed pulse-phase resolved spectroscopy in four time intervals during the MH state and found that \(E_\mathrm{cut}\) varies significantly with pulse phase: around the central hard peak (pencil beam) it reaches \(\sim 24\)–25 keV, while near the two shoulders (fan beam) it decreases to \(\sim 18\)–22 keV. Multiple RXTE observations of Her X-1/HZ Her between 1996 and 2005 also show that \(E_\mathrm{cut}\) decreases over the 35-day phase from \(\sim 21\ \mathrm{keV}\) to \(\sim 18\ \mathrm{keV}\) during the MH state \citep{leahy2022spectral}, which is consistent with our results.
\par
\par
In accreting X-ray pulsars, Fe K$\alpha$ lines are typically produced via fluorescence of material irradiated by the central source. When the iron line is produced via reflection from irradiated material, it usually originates from highly ionized regions, resulting in higher energies; lines produced via transmission are more likely from cooler material, showing lower fluorescence energies \citep{makishima1999cyclotron,leahy1999monte}. During the MH state of Her X-1, the Fe K$\alpha$ emission may originate from reprocessed emission in the magnetosphere or corona, from reflection off the far side of the inner accretion disk ring, and from transmission/reprocessing on the near side of the ring. As the 35-day phase evolves, the relative contributions from these regions change, leading to an evolution of the observed line energy. However, in this work, the Fe K$\alpha$ line energy does not show significant variations, likely due to the limited low-energy resolution of Insight-HXMT, which makes it difficult to measure small spectral shifts accurately.
\par
In the same period of Insight-HXMT observations, the average CRSF energy near the Main-On peak is \(\sim 38.5\ \mathrm{keV}\), slightly higher than \(\sim 38.2\ \mathrm{keV}\) reported by \citet{xiao2024insight}, but consistent with each other within the uncertainties. This minor difference is likely due to different energy ranges used in the fits, extending the fitting range up to 60 keV allows more complete sampling of high-energy photons, which affects constraints on the CRSF width and depth, and thus the determination of its centroid energy. Long-term observations indicate that the CRSF energy of Her X-1 decreased by \(\sim 4\ \mathrm{keV}\) between 1996 and 2012, then stabilized after 2012\citep{staubert2014long,ji2019long}. This change is generally not attributed to a decay of the neutron star’s global magnetic field, but rather to variations in the height of the CRSF-forming region in the accretion column or evolution of the magnetic field configuration near the polar caps. Within the Main High state of the 35-day cycle covered by our observations, the centroid energy of the CRSF remains nearly constant, further confirms that the CRSF energy of Her X-1 has remained stable since 2012. Although long-term observations indicate that Her X-1 is in the subcritical accretion regime \citep{staubert2007discovery,xiao2024insight}, where the CRSF energy is expected to increase with the X-ray flux in the main-high state, the precession of the accretion disk primarily changes the viewing geometry rather than the intrinsic accretion rate. As a result, the radiation height within the accretion column is not expected to vary significantly during the 35-day cycle. Previous studies have also not found a clear relation between the magnetic field strength at the emission region and the superorbital phase \citep{staubert2014long}.
\par

\section{Summary}
In this paper, we present a systematic study of the orbital dynamics and spectral properties of Her X-1 using Insight-HXMT observations during the 505th superorbital cycle. The timing analysis obtains new measurements of $T_{ecl}$ through pulse arrival time analysis and updates the orbital ephemeris of Her X-1 by combining these results with historical observations spanning approximately 40 years. Based on a quadratic ephemeris model, we re-estimated the orbital period derivative of the system, obtaining $\dot{P}_{\rm orb} = (-1.957 \pm 0.335) \times 10^{-11}~{\rm d\,d^{-1}}$. 
The new Insight-HXMT observations extend the temporal baseline for measuring the orbital parameters. The results indicate that the absolute value of the orbital period derivative is slightly lower than in previous studies but remains within the theoretically expected range. The evolution of the main spectral parameters during the Main High state are studied. The hydrogen column density $N_{\rm H}$ shows an overall increasing trend during the cycle. Specifically, we detected a CRSF at $\sim 38$ keV during the MH state, consistent with recent measurements. Combining with historical observations, our analysis further supports the conclusion that the CRSF energy of Her X-1 has remained generally stable since 2012.

\section*{Acknowledgements}
We are grateful to the referee for the suggestions to improve the manuscript. This work is supported by  the NSFC (12133007) and National Key Research and Development Program of China (Grants No. 2021YFA0718503 and 2023YFA1607901).

\bibliography{sample7}{}
\bibliographystyle{aasjournalv7}

\end{document}